\documentclass[letterpaper]{IEEEtran}

\usepackage[numbers,sort&compress]{natbib}

\makeatletter

\usepackage{etex}

\usepackage{zref-savepos}

\newcounter{mnote}

\def\xmarginnote{
  \xymarginnote{\hskip -\marginparsep \hskip -\marginparwidth}}

\def\ymarginnote{
  \xymarginnote{\hskip\columnwidth \hskip\marginparsep}}

\long\def\xymarginnote#1#2{
\vadjust{#1
\smash{\hbox{{
        \hsize\marginparwidth
        \@parboxrestore
        \@marginparreset
\footnotesize #2}}}}}

\def\mnoteson{
\gdef\mnote##1{\refstepcounter{mnote}\label{##1}
  \zsavepos{##1}
  \ifnum20432158>\number\zposx{##1}
  \xmarginnote{{\color{blue}\bf $\langle$\arabic{mnote}$\rangle$}}
  \else
  \ymarginnote{{\color{blue}\bf $\langle$\arabic{mnote}$\rangle$}}
  \fi
}
  }
\gdef\mnotesoff{\gdef\mnote##1{}}
\mnoteson
\mnotesoff

\newcommand{\figref}[1]{Fig.~\ref{#1}}

\usepackage{framed}

\usepackage{subcaption}
\usepackage{comment}

\usepackage[svgnames,dvipsnames]{xcolor}

\usepackage[all]{xy}

\usepackage{tikz}
\usetikzlibrary{positioning,matrix,through,calc,arrows,fit,shapes,decorations.pathreplacing,decorations.markings,}

\tikzstyle{block} = [draw,fill=blue!20,minimum size=2em]

\usepackage{qsymbols,amssymb,mathrsfs}
\usepackage[standard,thmmarks]{ntheorem}
\theoremstyle{plain}

\theoremsymbol{\ensuremath{_\vartriangleleft}}
\theorembodyfont{\itshape}
\theoremheaderfont{\normalfont\bfseries}
\theoremseparator{}

\theoremstyle{nonumberplain}
\theoremheaderfont{\scshape}
\theorembodyfont{\normalfont}
\theoremsymbol{\ensuremath{_\blacktriangleleft}}

\theoremnumbering{arabic}
\theoremstyle{plain}
\usepackage{latexsym}
\theoremsymbol{\ensuremath{_\Box}}
\theorembodyfont{\itshape}
\theoremheaderfont{\normalfont\bfseries}
\theoremseparator{}

\theorembodyfont{\upshape}
\theoremprework{\bigskip\hrule}
\theorempostwork{\hrule\bigskip}

\usepackage[overload]{empheq}

\let\iftwocolumn\if@twocolumn
\g@addto@macro\@twocolumntrue{\let\iftwocolumn\if@twocolumn}
\g@addto@macro\@twocolumnfalse{\let\iftwocolumn\if@twocolumn}

\mathtoolsset{showonlyrefs=false,showmanualtags}

\renewcommand{\eqref}[1]{\textup{(\refeq{#1})}} 
                                       
\newtagform{brackets}[\textbf]{(}{)}   
\usetagform{brackets}

\usepackage[Smaller]{cancel}

\PassOptionsToPackage{breaklinks,letterpaper,hyperindex=true,backref=false,bookmarksnumbered,bookmarksopen,linktocpage,colorlinks,linkcolor=BrickRed,citecolor=OliveGreen,urlcolor=Blue,pdfstartview=FitH}{hyperref}
\usepackage{hyperref}

\usepackage{cleveref}
\crefformat{footnote}{#2\footnotemark[#1]#3}

\usepackage{graphicx,psfrag}
\graphicspath{{figure/}{image/}}

\usepackage{diagbox}

\usepackage[ruled,vlined,linesnumbered]{algorithm2e}
\usepackage{algpseudocode}
\usepackage{listings} 
\lstdefinelanguage{Maple}{
  morekeywords={proc,module,end, for,from,to,by,while,in,do,od
    ,if,elif,else,then,fi ,use,try,catch,finally}, sensitive,
  morecomment=[l]\#,
  morestring=[b]",morestring=[b]`}[keywords,comments,strings]
\DeclareMathAlphabet{\mathpzc}{OT1}{pzc}{m}{it}
\usepackage{upgreek} 

\usepackage{dsfont}

\def\multi@nostar#1#2{
  \expandafter\def\csname multi#1\endcsname##1{
    \if ##1.\let\next=\relax \else
    \def\next{\csname multi#1\endcsname}     
    
    \expandafter\newcommand\csname #1##1\endcsname{#2}
    \fi\next}}

\def\multi@star#1#2{
  \expandafter\def\csname #1\endcsname##1{#2}
  \multi@nostar{#1}{#2}
}

\newcommand{\multi}{
  \@ifstar \multi@star \multi@nostar}

\multi*{rm}{\mathrm{#1}}
\multi*{mc}{\mathcal{#1}}
\multi*{op}{\mathop {\operator@font #1}}

\multi*{ds}{\mathds{#1}}
\multi*{set}{\mathcal{#1}}
\multi*{rsfs}{\mathscr{#1}}
\multi*{pz}{\mathpzc{#1}}
\multi*{M}{\boldsymbol{#1}}
\multi*{R}{\mathsf{#1}}
\multi*{RM}{\M{\R{#1}}}
\multi*{bb}{\mathbb{#1}}
\multi*{td}{\tilde{#1}}
\multi*{tR}{\tilde{\mathsf{#1}}}
\multi*{trM}{\tilde{\M{\R{#1}}}}
\multi*{tset}{\tilde{\mathcal{#1}}}
\multi*{tM}{\tilde{\M{#1}}}
\multi*{baM}{\bar{\M{#1}}}
\multi*{ol}{\overline{#1}}

\multirm  ABCDEFGHIJKLMNOPQRSTUVWXYZabcdefghijklmnopqrstuvwxyz.
\multiol  ABCDEFGHIJKLMNOPQRSTUVWXYZabcdefghijklmnopqrstuvwxyz.
\multitR   ABCDEFGHIJKLMNOPQRSTUVWXYZabcdefghijklmnopqrstuvwxyz.
\multitd   ABCDEFGHIJKLMNOPQRSTUVWXYZabcdefghijklmnopqrstuvwxyz.
\multitset ABCDEFGHIJKLMNOPQRSTUVWXYZabcdefghijklmnopqrstuvwxyz.
\multitM   ABCDEFGHIJKLMNOPQRSTUVWXYZabcdefghijklmnopqrstuvwxyz.
\multibaM   ABCDEFGHIJKLMNOPQRSTUVWXYZabcdefghijklmnopqrstuvwxyz.
\multitrM   ABCDEFGHIJKLMNOPQRSTUVWXYZabcdefghijklmnopqrstuvwxyz.
\multimc   ABCDEFGHIJKLMNOPQRSTUVWXYZabcdefghijklmnopqrstuvwxyz.
\multiop   ABCDEFGHIJKLMNOPQRSTUVWXYZabcdefghijklmnopqrstuvwxyz.
\multids   ABCDEFGHIJKLMNOPQRSTUVWXYZabcdefghijklmnopqrstuvwxyz.
\multiset  ABCDEFGHIJKLMNOPQRSTUVWXYZabcdefghijklmnopqrstuvwxyz.
\multirsfs ABCDEFGHIJKLMNOPQRSTUVWXYZabcdefghijklmnopqrstuvwxyz.
\multipz   ABCDEFGHIJKLMNOPQRSTUVWXYZabcdefghijklmnopqrstuvwxyz.
\multiM    ABCDEFGHIJKLMNOPQRSTUVWXYZabcdefghijklmnopqrstuvwxyz.
\multiR    ABCDEFGHIJKL NO QRSTUVWXYZabcd fghijklmnopqrstuvwxyz.
\multibb   ABCDEFGHIJKLMNOPQRSTUVWXYZabcdefghijklmnopqrstuvwxyz.
\multiRM   ABCDEFGHIJKLMNOPQRSTUVWXYZabcdefghijklmnopqrstuvwxyz.

\newcommand{\dotleq}{\buildrel \textstyle  .\over {\smash{\lower
      .2ex\hbox{\ensuremath\leqslant}}\vphantom{=}}}
\newcommand{\dotgeq}{\buildrel \textstyle  .\over {\smash{\lower
      .2ex\hbox{\ensuremath\geqslant}}\vphantom{=}}}

\newcommand{\bM}{\begin{bmatrix}}
\newcommand{\eM}{\end{bmatrix}}
\newcommand{\bSM}{\left[\begin{smallmatrix}}
\newcommand{\eSM}{\end{smallmatrix}\right]}
\renewcommand*\env@matrix[1][*\c@MaxMatrixCols c]{
  \hskip -\arraycolsep
  \let\@ifnextchar\new@ifnextchar
  \array{#1}}

\newqsymbol{`N}{\mathbb{N}}
\newqsymbol{`R}{\mathbb{R}}
\newqsymbol{`P}{\mathbb{P}}
\newqsymbol{`Z}{\mathbb{Z}}

\newqsymbol{`|}{\mid}

\newqsymbol{`8}{\infty}
\newqsymbol{`1}{\left}
\newqsymbol{`2}{\right}
\newqsymbol{`6}{\partial}
\newqsymbol{`0}{\emptyset}
\newqsymbol{`-}{\leftrightarrow}
\newqsymbol{`<}{\langle}
\newqsymbol{`>}{\rangle}

\DeclarePairedDelimiter\abs{\lvert}{\rvert}

\DeclarePairedDelimiter\Set{\{}{\}}
\newcommand{\imod}[1]{\allowbreak\mkern10mu({\operator@font mod}\,\,#1)}

\newcommand{\threecols}[3]{
\hbox to \textwidth{
      \normalfont\rlap{\parbox[b]{\textwidth}{\raggedright#1\strut}}
        \hss\parbox[b]{\textwidth}{\centering#2\strut}\hss
        \llap{\parbox[b]{\textwidth}{\raggedleft#3\strut}}
    }
}

\newcommand{\reason}[2][\relax]{
  \ifthenelse{\equal{#1}{\relax}}{
    \left(\text{#2}\right)
  }{
    \left(\parbox{#1}{\raggedright #2}\right)
  }
}

\let\SavedDoubleVert\relax
\let\protect\relax
{\catcode`\|=\active
  \xdef\extendvert{\protect\expandafter\noexpand\csname extendvert \endcsname}
  \expandafter\gdef\csname extendvert \endcsname#1{\mskip-5mu \left.
      \ifx\SavedDoubleVert\relax \let\SavedDoubleVert\|\fi
     \:{\let\|\SetDoubleVert
       \mathcode`\|32768\let|\SetVert
     #1}\:\right.\mskip-5mu}
}
\def\SetVert{\@ifnextchar|{\|\@gobble}
    {\egroup\;\mid@vertical\;\bgroup}}
\def\SetDoubleVert{\egroup\;\mid@dblvertical\;\bgroup}

\begingroup
 \edef\@tempa{\meaning\middle}
 \edef\@tempb{\string\middle}
\expandafter \endgroup \ifx\@tempa\@tempb
 \def\mid@vertical{\middle|}
 \def\mid@dblvertical{\middle\SavedDoubleVert}
\else
 \def\mid@vertical{\mskip1mu\vrule\mskip1mu}
 \def\mid@dblvertical{\mskip1mu\vrule\mskip2.5mu\vrule\mskip1mu}
\fi

\makeatother

\usepackage{fouridx}

\usepackage{framed}
\usetikzlibrary{positioning,matrix,decorations.shapes}

\usepackage{paralist}

\usepackage{enumerate}

\usepackage[normalem]{ulem}

\numberwithin{equation}{section}
\makeatletter
\@addtoreset{equation}{section}
\renewcommand{\theequation}{\arabic{section}.\arabic{equation}}
\@addtoreset{Theorem}{section}
\renewcommand{\theTheorem}{\arabic{section}.\arabic{Theorem}}
\@addtoreset{Lemma}{section}
\renewcommand{\theLemma}{\arabic{section}.\arabic{Lemma}}
\@addtoreset{Corollary}{section}
\renewcommand{\theCorollary}{\arabic{section}.\arabic{Corollary}}
\@addtoreset{Example}{section}
\renewcommand{\theExample}{\arabic{section}.\arabic{Example}}
\@addtoreset{Remark}{section}
\renewcommand{\theRemark}{\arabic{section}.\arabic{Remark}}
\@addtoreset{Proposition}{section}
\renewcommand{\theProposition}{\arabic{section}.\arabic{Proposition}}
\@addtoreset{Definition}{section}
\renewcommand{\theDefinition}{\arabic{section}.\arabic{Definition}}
\@addtoreset{Subclaim}{Theorem}
\renewcommand{\theSubclaim}{\theTheorem\Alph{Subclaim}}
\makeatother

\newenvironment{ybox}{
  \setlength{\FrameSep}{1mm}
  \setlength{\FrameRule}{0mm}
  
  \MakeFramed {\FrameRestore}}
{\endMakeFramed}

\newenvironment{gbox}{
	  \setlength{\FrameSep}{1mm}
	  \setlength{\FrameRule}{0mm}
  
  \MakeFramed {\FrameRestore}}
{\endMakeFramed}

\newcommand{\pbox}[2][\relax]{
  \setlength{\FrameSep}{1.5mm}
 \setlength{\FrameRule}{0mm}
  \begin{gbox}
    \noindent\underline{#1:}\newline
    #2
  \end{gbox}
}

\title{Incremental and Decremental Secret Key Agreement}

\author{Chung Chan, Ali Al-Bashabsheh and Qiaoqiao Zhou
	\thanks{C.\ Chan (email: cchan@inc.cuhk.edu.hk),
		A.\ Al-Bashabsheh, and Q.\ Zhou are with the Institute of Network Coding at the
		Chinese University of Hong Kong, the Shenzhen Key Laboratory of
		Network Coding Key Technology and Application, China, and the
		Shenzhen Research Institute of the Chinese University of Hong
		Kong.
	}
	\thanks{The work is supported in part by the grants from the University Grants Committee of the Hong Kong SAR, China (Project
		No. AoE/E-02/08 and 14200714), Shenzhen Research Fund (KQCX20130628164008004) and Shenzhen Key Laboratory of Network Coding
		Key Technology and Application, Shenzhen, China
		(ZDSY20120619151314964).}
	}

\begin{document}

\IEEEoverridecommandlockouts
\maketitle

\begin{abstract}

We study the rate of change of the multivariate mutual information among a set of random variables when some common randomness is added to or removed from a subset. This is formulated more precisely as two new multiterminal secret key agreement problems which ask how one can increase the secrecy capacity efficiently by adding common randomness to a small subset of users, and how one can simplify the source model by removing redundant common randomness that does not contribute to the secrecy capacity. 
The combinatorial structure has been clarified along with some meaningful open problems.
\end{abstract} 


\section{Introduction}
\label{sec:introduction}

We consider the multiterminal secret key agreement problem formulated by \cite{csiszar04}. A group of users, each observing a private correlated random source, discuss
in public so they can agree on a secret key. The key is a random variable that needs to be recoverable by
every user after the discussion. Furthermore, the key must be secured against a wiretapper who can
observe the entire discussion among the users but has no access to their private sources. 

The maximum achievable secret key rate is called the \emph{secrecy capacity}. It was characterized by
\cite{csiszar04} as a linear program. Because the wiretapper can listen to the entire public
discussion, \emph{the randomness of the secret key can only come from the information mutual to the
private correlated source components.}
Indeed, in the two-user case, it was shown in \cite{csiszar04} that the capacity is equal to
\emph{Shannon's mutual information}: 
\begin{align}
	I(\RZ_1\wedge \RZ_2) &= D(P_{\RZ_1\RZ_2}\|P_{\RZ_1}P_{\RZ_2}),\label{eq:I}
\end{align}
where each user $i\in \Set{1,2}$ observes the discrete memoryless correlated private source $\RZ_i$. The mutual information above is written as the \emph{divergence} $D$ from the joint distribution $P_{\RZ_1\RZ_2}$ to the product of the marginal distributions $P_{\RZ_1}$ and $P_{\RZ_2}$. 

In the multiterminal case, let $V$ be the set of (two or more) users, and $\RZ_V:=(\RZ_i\mid i\in V)$ be a random vector where $\RZ_i$ is a discrete memoryless source component privately observed by user $i\in V$. There was a divergence upper bound on the capacity in~\cite{csiszar04}, which was identified~\cite{chan2008tightness,chan10md} to be tight in the special case without helpers, giving rise to the alternative capacity characterization:
\begin{subequations}
\begin{align}
	I(\RZ_V) &:= \min_{\mcP\in \Pi'(V)} I_{\mcP}(\RZ_V), \kern2em \text{where}\label{eq:MMI}\\
	I_{\mcP}(\RZ_V)&:=\frac{D(P_{\RZ_V}\|\prod_{C\in \mcP} P_{\RZ_C})}{\abs {\mcP}-1},\label{eq:IP}
\end{align}
\end{subequations}
and $\Pi'(V)$ is the collection of partitions $\mcP$ of $V$ into two or more non-empty disjoint sets.

Following \cite{chan15mi},
we call \eqref{eq:MMI} the \emph{multivariate mutual information (MMI)}. It is easy
to see that \eqref{eq:I} is a special case of \eqref{eq:MMI} when $V=\Set{1,2}$. Indeed, the MMI was
formally regarded in \cite{chan15mi} as a measure of mutual information among multiple random
variables, thereby extending various interpretations and properties of Shannon's mutual information to the multivariate case. The MMI has other operational meanings, ranging from
tree-packing~\cite{nitinawarat10}, hypergraph connectivity~\cite{chan10md}, undirected network
coding~\cite{chan11isit},  vocality constraints~\cite{mukherjee14,mukherjee15,zhang15}, successive omniscience~\cite{chan16so,ding15} and data
clustering~\cite{chan15allerton,chan16cluster}.

In this work, we want to study how the MMI of a set of random variables changes with respect to
changes in the MMI of a subset of the random variables. We formulate two new problems, called the
\emph{incremental secret key agreement (ISKA)} and \emph{decremental secret key agreement (DSKA)}.
In ISKA, a subset of users is given an additional common randomness in the form of a random source of
certain entropy rate. The objective is to find an efficient resource allocation, i.e., to increase
the capacity as much as possible without requiring too much common randomness to be added to too
many users. In DSKA, we remove some common randomness from a subset of users. The objective is to simplify the source model, but without reducing the capacity much.
In particular, we want to identify redundant common randomness whose removal does not diminish the capacity.\footnote{The idea of redundant common randomness first appeared in \cite[Proposition~2.3]{chan10md}. It was called excess edge under a hypergraphical source model, and was related to the notion of partition connectivity for hypergraphs. The idea was also used in \cite{courtade16} in characterizing the minimum communication complexity for secret key agreement, but under a non-asymptotic hypergraphical source model when the communication protocol is restricted to be linear. \cite{MKS16,mukherjee16,chan16itw} further considered the non-asymptotic case and derived more efficiently computable bounds on the communication complexity.} 

\section{Motivation}
\label{sec:motivation}

We first explain the idea using a simple example. Define the random source as
\begin{align*}
	\RZ_1 := (\RX_a,\RX_b),\kern1em	\RZ_2 := (\RX_a,\RX_b),\text{ and }
	\RZ_3 := \RX_a,
\end{align*}
where $\RX_a$ and $\RX_b$ are independent uniformly random bits. The random bits $\RX_j$'s determine
the correlation, or joint distribution, of the sources $\RZ_i$'s. Let $V:=\Set{1,2,3}$ be the set of
users. Each user $i\in V$ observes the discrete memoryless source $\RZ_i$ privately. It is easy to
see that the users can agree on a secret key bit, namely, $\RX_a$, without any public discussion.
In fact, the users cannot agree on any more secret key bits, even with additional public discussion.
This is clear since $\RX_a$ is already the entire private observation of user $3$. \emph{The secrecy capacity is
	therefore $1$~bit.}

For ISKA, we consider adding a common randomness to a subset of users. For example, we may add a new
independent bit $\RX_c$ to the private sources of users~$2$ and $3$ as follows:
\begin{align*}
	\RZ'_1 := (\RX_a,\RX_b),\kern1em
	\RZ'_2 := (\RX_a,\RX_b,\RX_c),\text{ and }
	\RZ'_3 := (\RX_a,\RX_c),
\end{align*}  
where user $i\in V$ observes the new source $\RZ_i'$ instead of $\RZ_i$.
With such an increment to the private sources, the bit $\RX_b$ can also be used as a secret key in
addition to $\RX_a$, i.e., the secret key can be chosen as $\RK=(\RX_a,\RX_b)$. To achieve this, user~$2$ can reveal
in public the XOR $\RF:=\RX_b\oplus \RX_c$, and so user $3$ can recover $\RX_b$ by subtracting
$\RX_c$ from the sum. It can be shown that $\RK$ is independent of $\RF$ and is therefore kept
secret from a wiretapper observing the entire public discussion. \emph{The secrecy capacity is now equal
	to $2$~bits.} 

In the above, the addition of the private common randomness $\RX_c$ increased the secrecy capacity by $1$ bit.
If we are allowed to choose who to give this common randomness to,
the current choice of users $2$ and
$3$ is in fact the most efficient (besides the equivalent choice of users~1 and 3).
For example, if $\RX_c$ were given to users~$1$ and $2$
instead, then it is evident that the capacity would not have increased.
%
Of course, one may choose to give $\RX_c$  to
user~$1$ in addition to users $2$ and $3$,
where the capacity would be 2 bits.
However,
such an allocation is not considered efficient since additional resources, e.g.,  private communication, may be needed to give $\RX_c$ to user~$1$.

For DSKA, we consider removing  some common randomness from a subset of users, while trying not to
diminish the secrecy capacity. It is easy to see that removing $\RX_b$ from users~$1$ and $2$ does
not diminish the capacity, while removing $\RX_a$ from all the users does.\footnote{Note that we	remove a common randomness from all the users observing it.
The source is simpler after such a removal since its joint entropy reduces strictly by the amount of
common randomness we remove. It is possible to extend this to more general multi-letter preprocessing of the random source, and doing so is potentially more useful for non-hypergraphical sources.} 
In other words, the common randomness $\RX_b$ is redundant but $\RX_a$ is not. We can
therefore consider the following simpler source for the purpose of achieving the secrecy capacity of
$1$~bit:
\begin{align*}
	\RZ''_i&:=\RX_a &&\text{for $i\in \Set{1,2,3}$.}
\end{align*}
Simplifying the source is useful because it simplifies the capacity-achieving scheme in \cite{csiszar04} by reducing the amount of discussion required for the \emph{communication for omniscience}.

\section{Problem formulation}

We will formulate ISKA and DSKA as extensions of the secret key agreement problem in
\cite{csiszar04} under the source model without wiretapper's side information nor helpers. Readers
may refer to \cite{csiszar04} for the detailed secret key agreement protocol. In our formulation, we
will only need the characterization~\eqref{eq:MMI}~\cite{chan10md} of the secrecy capacity. In this paper, we primarily 
denote sets with capital letters, random variables with the sans-serif font, and families of sets with
script typeface letters. Furthermore, for a family $\mcF$ of sets, we will use $\op{minimal} \mcF$ and
$\op {maximal} \mcF$ to denote, respectively, the sets of inclusion-wise minimal and maximal
elements of $\mcF$.

\subsection{Incremental secret key agreement}

To formulate ISKA, we consider adding a common randomness $\RX$ of entropy $`e>0$ to a subset $S$ of users: 
\begin{Definition}
	For $S\subseteq V$ and $`e>0$, we say that $\RZ_V^{S,`e}$ is an $(S,`e)$-incremented source of $\RZ_V$ if it can be written as
	\begin{align}
		\RZ_i^{S,`e}:=\begin{cases}
			(\RZ_i,\RX) & i\in S\\
			\RZ_i & \text{otherwise,}
		\end{cases}
	\end{align}
	where $\RX$ is independent of $\RZ_V$ and has entropy $H(\RX)=`e$.
\end{Definition}
We want to characterize the \emph{rate of increase} in the secrecy capacity~\eqref{eq:MMI} of the incremented source:
\begin{Definition}
	The subderivative (one-sided limit)
	\begin{align}
		`r_{S}^+(\RZ_V) &:= \left.\tfrac{\partial I(\RZ_V^{S,`e})}{\partial `e}`2|_{`e=0^+}  = \lim_{`e\downarrow 0} \frac{I(\RZ_V^{S,`e})-I(\RZ_V)}{`e}\label{eq:`r_S^+}
	\end{align}
	is the \emph{growth rate} of the secrecy capacity for the private source $\RZ_V$ with respect to an infinitesimal increment in the common randomness of the subset $S$. Since the additional common randomness $\RX$ is a valuable resource, we want to maximize the growth rate among all subsets $S$ of the same size as in
	\begin{align}
		`r_k^+(\RZ_V) &:= \max_{S\subseteq V:\abs{S}\leq k} `r_{S}^+(\RZ_V) \label{eq:`r_k^+}
	\end{align} 
	for integer $k$. We refer to $`r_k^+$ as the \emph{growth rate of order} $k$. 
\end{Definition}

For an efficient allocation of common randomness, we want to identify small subsets with strictly positive growth rate:
\begin{Definition}
	\label{def:critical}
	$S\subseteq V$ is said to be a \emph{critical (hyper)edge} if it is inclusion-wise \emph{minimal} with $`r_S^+(\RZ_V)>0$. We use
	\begin{align}
		\pzS_{\op{crit}}(\RZ_V) &:=\op{minimal}\Set*{S\subseteq V\mid `r_S^+(\RZ_V)>0}
	\end{align}
	to denote the set of all critical edges.\footnote{Unlike \cite{courtade16}, the word critical is associated with an edge rather than a set family for a non-asymptotic hypergraphical source. Furthermore, the notion of critical family in \cite{courtade16} is related to the notion of excess edge in DSKA rather than the notion of critical edge in the ISKA problem. We also consider an asymptotic source model that is not restricted to be hypergraphical.}
\end{Definition}
(We remark that a critical edge $S$ is an edge in the
$(S,\epsilon)$-incremented source.)
Out of all the critical edges, the subsets with minimum size require the least resource. It is easy to argue that the minimum critical edges are the optimal solutions to \eqref{eq:`r_k^+} for the smallest $k$ such that $`r_k^+(\RZ_V)>0$.

\subsection{Decremental secret key agreement}
To formulate the DSKA problem, we will consider a special kind of random sources: 
\begin{Definition}
	We say that a source $\RZ_V$ has an edge $S\subseteq V$ if there is a common randomness $\RX'$
	that is observed only by the users in $S$,
	i.e., if we can rewrite $\RZ_i$ (up to bijection) as
	\begin{align}
		\RZ_i = \begin{cases}
			(\RZ'_i,\RX') & i\in S\\
			\RZ'_i & \text{otherwise},
		\end{cases}
	\end{align}
	where $\RZ'_V$ is independent of $\RX'$. We reduce  such a source to the following $(S,`e)$-decremented source by removing an $`e\in (0, H(\RX')]$ amount of common randomness $\RX'$:
	\begin{align}
		\RZ_i^{S,-`e} &= \begin{cases}
			(\RZ'_i,\RX) & i\in S\\
			\RZ'_i
		\end{cases}
	\end{align} 
	for some common randomness $\RX$ independent of $\RZ_V'$ and with $H(\RX)=H(\RX')-`e$.
\end{Definition}

Contrary to ISKA, we are interested in the \emph{rate of decrease} in the secrecy capacity~\eqref{eq:MMI}: 
\begin{Definition}
	The subderivative (one-sided limit)
	\begin{align}
		\kern-1em `r_{S}^-(\RZ_V) &:= \left.-\tfrac{\partial I(\RZ_V^{S,-`e})}{\partial `e}`2|_{`e=0^+}  \kern-1em= \lim_{`e\downarrow 0} \frac{I(\RZ_V)-I(\RZ_V ^{S,-`e})}{`e},\kern-.5em
	\end{align}
	is the \emph{loss rate} of the secrecy capacity for the private source $\RZ_V$ with edge $S$. 
\end{Definition}
Unlike ISKA, we are interested in edges $S$ with zero loss rate. 

\begin{Definition}
	$S\subseteq V$ is said to be an \emph{excess} or \emph{redundant} edge if the corresponding loss
	rate is $0$ for the source $\RZ_V$ with edge $S$. In this case, we can simplify the secret key
	agreement by removing common randomness of the edge $S$ without diminishing the secrecy capacity.  
\end{Definition}

\section{Main results}

As pointed out by \cite{chan11isit,milosavljevic11} and elaborated in \cite{chan15mi}, the MMI can be computed in polynomial time using 
 \emph{submodular function
	minimization}~\cite{schrijver02} algorithms. 
A
polynomial time algorithm was also given by an earlier work of
Fujishige~\cite{fujishige84,fujishige88} for a more general type of \emph{submodular functions}
and \emph{set family}.
It turns out that the characterization and computation of the growth rate, loss rate, critical
edges, and excess edges depend only on the optimal partitions that attain the MMI~\eqref{eq:MMI}. The set of the optimal partitions will be denoted by
\begin{align}
	\Uppi^*(\RZ_V) :=\Set*{\mcP\in \Pi'(V)\mid I_{\mcP}(\RZ_V)=I(\RZ_V)}.
\end{align}
Using the combinatorial result of \cite{narayanan90}, the set $\Uppi^*(\RZ_V)\cup \Set{V}$ 
forms a lattice, and hence admits a unique finest partitionm 
\begin{align}
	\pzP^*(\RZ_V):= \min \Uppi^*(\RZ_V)\label{P^*}.
\end{align}
Here, the minimum is with respect to the partial order ``$\prec$'' of the partitions, which is defined
as $\mcP\prec \mcP'$ if the partition $\mcP$ \emph{is finer than} (or \emph{a refinement of}) the
partition $\mcP'$. In other words, $\mcP$ can be obtained by further partitioning one or more
subsets in $\mcP'$. Following \cite{chan15mi}, we will refer to the finest partition $\pzP^*(\RZ_V)$
as the \emph{fundamental partition}.
This partition has an elegant interpretation~\cite{chan15allerton} in
data clustering, and furthermore, can be computed in strongly polynomial time using algorithms
such as \cite{nagano10} applied to the \emph{minimum average cost clustering}.

For ISKA, we can characterize the growth rate and critical edges using the optimal partitions as follows:
\begin{Theorem}
	\label{thm:rhopatch}
	\begin{subequations}
		For any $\RZ_V$ and $S\subseteq V$, 
		\begin{align}
			`r^+_{S}(\RZ_V) &= \min_{\mcP \in \Uppi^*(\RZ_V)} \frac{\sum_{C\in
					\mcP} `c_{\Set{C \cap S\neq `0}} - `c_{\Set{S\neq `0}}
			}{\abs{\mcP}-1}, \label{eq:rhoS}
		\end{align}
		where $`c$ is the indicator function of the condition specified in the subscript.
		It follows that
		\begin{align}
			\pzS_{\op{crit}}(\RZ_V) &= \op{minimal}\{S\subseteq V\mid \notag \\ 
			& \kern4em S\not\subseteq C,\forall C\in
			\mcP\in \Uppi^*(\RZ_V)\}. \label{eq:patch}
		\end{align}
		In other words, $S\subseteq V$ is critical iff it is a minimal set that overlaps at least two blocks of every optimal partition.
	\end{subequations}
\end{Theorem}
\begin{Proof}
	See Appendix~\ref{sec:A}.
\end{Proof}
Indeed, $\pzS_{\op{crit}}(\RZ_V)$ depends on $\RZ_V$ only through the coarsest optimal
partitions in $\Uppi^*(\RZ_V)$. This is because, if $S$ crosses a partition, i.e., overlaps at least
two blocks of the partition, then it must also cross any refinement of the partition.

For DSKA, we can similarly characterize the loss rate and excess edges as follows:
\begin{Theorem}
	\label{thm:rho-}
	\begin{subequations}
		For any $\RZ_V$ with edge $S\subseteq V$, 
		\begin{align}
			`r^-_{S}(\RZ_V) &= \max_{\mcP \in \Uppi^*(\RZ_V)} \frac{\sum_{C\in
					\mcP} `c_{\Set{C \cap S\neq `0}} - `c_{\Set{S\neq `0}}
			}{\abs{\mcP}-1}. \label{eq:rhoS-}
		\end{align}
		It follows that $S$ is an excess edge iff 
		\begin{align}
			\exists C\in \pzP^*(\RZ_V), S\subseteq C,\label{eq:excess}
		\end{align}
		i.e., $S$ does not cross the fundamental partition.
	\end{subequations}
\end{Theorem}
\begin{Proof}
	See Appendix~\ref{sec:A}.
\end{Proof}
The condition for an excess edge depends only on the fundamental partition, and therefore can be
checked in strongly polynomial time.

Equations \eqref{eq:rhoS} and \eqref{eq:patch} imply the following simple properties of $`r^+_k$~\eqref{eq:`r_k^+} and $\pzS_{\op{crit}}$. 
\begin{Proposition}
	\label{pro:rhopatch}
	$`r_k^+(\RZ_V)$ is non-decreasing in $k$ and equal to $0$ for
	$k\leq 1$. Furthermore,
	\begin{align*}
		`r_k^+(\RZ_V) =1\kern1em \text{ iff }\kern1em k\geq \abs{\pzP^*(\RZ_V)},
	\end{align*}
	the size of the fundamental partition. Finally,  we have at least one critical edge,
	i.e. $\pzS_{\op{crit}}(\RZ_V)\neq `0$, and the minimum size of a critical edge is at least $2$,
	i.e., $\min\Set{\abs {S}\mid S\in \pzS_{\op {crit}}(\RZ_V)}\geq 2$.
\end{Proposition}
\begin{Proof}
	See Appendix~\ref{sec:A}.
\end{Proof}

Computing $`r_k^+$ in general can be quite difficult but some simple cases will be given in the next section. Surprisingly, it turns out that
computing a minimum critical edge can be done in strongly polynomial time, which  
is due to the result below.
\begin{Theorem}
	\label{thm:C}
	All critical edges in $\pzS_{\op{crit}}(\RZ_V)$ have the same size, and are therefore minimum.
\end{Theorem}
In other words, all the critical edges are minimum. A critical edge can be obtained easily as follows: Starting with $S=V$, repeatedly remove an element from $S$
as long as doing so does not violate
$I(\RZ_V^{S,1})>I(\RZ_V)$. The condition can be checked
in strongly polynomial time for at most $O(\abs{V}^2)$ times.

Indeed, a stronger result can be proved. We will characterize the
critical edges completely using only
the maximal blocks from the optimal partitions,
\begin{align}
	\label{eq:T}
	\pzT_{\max}(\RZ_V) &:= \op {maximal} \bigcup \Uppi^*(\RZ_V)
\end{align}
which can also be computed in strongly polynomial time. More
precisely, we can show that:
\begin{Lemma}
	\label{lem:T}
	Either one of the following cases happen:
	\begin{align}
		\pzT_{\max}(\RZ_V) &\in \Uppi^*(\RZ_V) \label{eq:T1}\\
		\bar{\pzT}_{\max}(\RZ_V) &:= \Set{V`/C:C\in
			\pzT_{\max}(\RZ_V)}\in \Pi'(U)\label{eq:T2}
	\end{align}
	for some $U\subseteq V$. In words, either $\pzT_{\max}(\RZ_V)$ is an
	optimal partition or its complement $\bar{\pzT}_{\max}(\RZ_V)$ is a
	set of at least two non-empty disjoint subsets of $V$. Indeed, \eqref{eq:T1} means that $\pzT_{\max}(\RZ_V)$ is the unique coarsest optimal partition.
\end{Lemma}
\begin{Theorem}
	\label{thm:TC}
	If \eqref{eq:T1} happens,
	\begin{align}
		\pzS_{\op{crit}}(\RZ_V) &= \Set{\Set{i,j}\mid i\in C,j\in V`/C,C\in
			\pzT_{\max}(\RZ_V)}\label{eq:C1}
	\end{align}
	and so all the critical edges have size $2$. 
	If \eqref{eq:T2} happens,
	\begin{align}
		\pzS_{\op{crit}}(\RZ_V) &= \Set{\Set{i_C\mid C\in \pzT_{\max}(\RZ_V)}\mid i_C\in V`/C},\label{eq:C2}
	\end{align}
	which is taken to mean the collection of the sets $\Set{i_C\mid C\in \pzT_{\max}(\RZ_V)}$ of representatives $i_C$ of subsets $V`/C$ for $C\in \pzT_{\max}(\RZ_V)$. It follows that $\abs{\pzT_{\max}(\RZ_V)}$ is the size of all the critical edges.
\end{Theorem}
Note that, Theorem~\ref{thm:TC} implies Theorem~\ref{thm:C} immediately.
\begin{Proof}
	See Appendix~\ref{sec:B}.
\end{Proof}

\begin{Example}
	\label{eg:patch}
	Consider $V=\Set{1,2,3}$. Let $\RZ_1=\RZ_2$ be a uniformly random bit, and $\RZ_3=0$. It can be shown that
	\begin{align*}
		\Uppi^*(\RZ_V) &= \Set{\Set{\Set{1,2},\Set{3}}}\\
		\pzS_{\op {crit}}(\RZ_V) &= \Set{\Set{1,3},\Set{2,3}}.
	\end{align*}
	There is a unique optimal partition and
	so $\pzT_{\max}(\RZ_V)=\pzP^*(\RZ_V)=\Set{\Set{1,2},\Set{3}}\in
	\Uppi^*(\RZ_V)$, satisfying \eqref{eq:T1}. The set of
	critical edges by \eqref{eq:C1} is
	$\pzS_{\op{crit}}(\RZ_V)=\Set{\Set{1,3},\Set{2,3}}$.
	
	Indeed, \eqref{eq:T1} may hold even when the optimal partition is
	not unique. For instance, consider $V=\Set{1,2,3}$ and let
	$\RZ_1\!:=\!(\RX_a,\RX_b,\RX_c)$, $\RZ_2\!:=\!(\RX_a,\RX_b,\RX_d)$ and $\RZ_3\!:=\!(\RX_c,\RX_d)$
	where $\RX_i$'s are uniformly random and
	independent bits. It follows that
	\begin{align*}
	\Uppi^*(\RZ_V) &=
	\Set{\Set{\Set{1,2},\Set{3}},\Set{\Set{1},\Set{2},\Set{3}}}\\
	\pzT_{\max}(\RZ_V) &= \Set{\Set{1,2},\Set{3}}\in \Uppi^*(\RZ_V)\\
	\pzS_{\op{crit}}(\RZ_V)&= \Set{\Set{1,3},\Set{2,3}}
	\end{align*}
	which satisfies \eqref{eq:T1} but $\pzT_{\max}$ is the coarsest partition
	rather than the
	fundamental partition.
\end{Example}

\begin{Example}
	\label{eg:tree}
	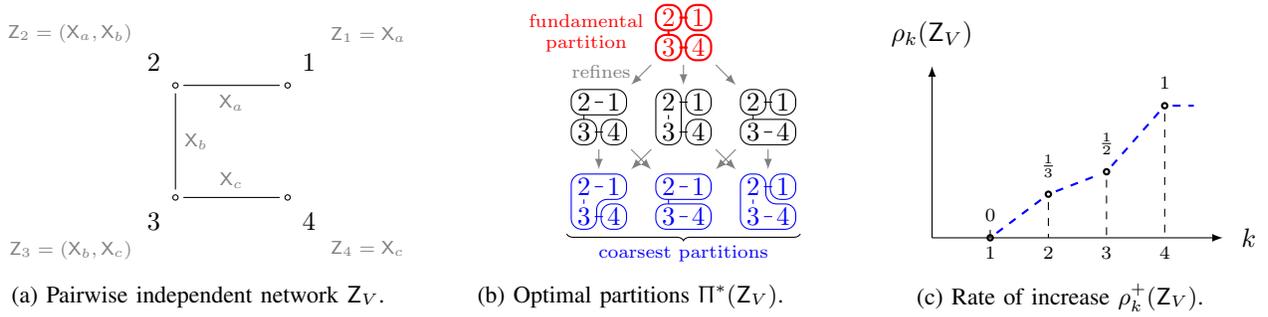
\begin{figure*}
		\centering
		\tikzstyle{point}=[draw,circle,minimum size=.2em,inner sep=0,
		outer sep=.2em]
		\subcaptionbox{Pairwise independent network $\RZ_V$.\label{fig:egrk:source}}{
			\begin{tikzpicture}[x=.6em,y=.6em,>=latex]
			\foreach \x/\angle/\lb in {1/45/{$\RZ_\x=\RX_a$},2/135/{$\RZ_\x=(\RX_a,\RX_b)$},3/-135/{$\RZ_\x=(\RX_b,\RX_c)$},4/-45/{$\RZ_\x=\RX_c$}}
			{
				\path (\angle:5) node (\x) [point,label=\angle:$\x$,label={[label distance=1.5em]\angle:{\color{gray}\scriptsize\lb}}] {};
			}
			\foreach \x/\y/\lp/\lb/\lpp/\lbb in {3/4/above/$\RX_c$/below/~,
				1/2/below/$\RX_a$/above/~,
				2/3/right/$\RX_b$/left/~}
			\draw[-] (\x) to node [label=\lp:{\color{gray}\scriptsize\lb},label=\lpp:{\color{blue}\scriptsize\lbb}]{} (\y);
			\end{tikzpicture}
		}
		\hfil
		\subcaptionbox{Optimal partitions $\Uppi^*(\RZ_V)$.\label{fig:egrk:optpart}}{
                  \includegraphics{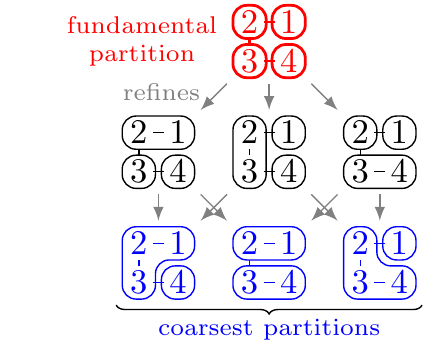}
		}
		\hfil
		\subcaptionbox{Rate of increase $`r_k^+(\RZ_V)$.\label{fig:egrk:`r_k}}{
			{\def\ux{2.2}
				\def\uy{5}
				\begin{tikzpicture}[x=1em,y=1em,>=latex]
				\draw[->] (0,0) -- (0,1.3*\uy) node [label=above:$`r_k(\RZ_V)$] {};
				\draw[->] (0,0) -- (5*\ux,0) node (k) [label=right:$k$] {};
				\path (1*\ux,0*\uy) node (1) [point,thick,label=above:{\scriptsize$0$}] {};
				\path (2*\ux,0.33*\uy) node (2) [point,thick,label=above:{\scriptsize$\frac13$}] {};
				\path (3*\ux,0.5*\uy) node (3) [point,thick,label=above:{\scriptsize$\frac12$}] {};
				\path (4*\ux,1*\uy) node (4) [point,thick,label=above:{\scriptsize$1$}] {};
				\draw[dashed,blue,thick]
				(1)--(2)--(3)--(4)--(4.5*\ux,1*\uy);
				\foreach \x in {1,2,3,4}
				\draw[dashed] (\x) -- (\x|-k) node [below] {\scriptsize $\x$};
				\end{tikzpicture}
			}
		}
		\caption{Optimal partitions and rate of increase of the tree
			network in Example~\ref{eg:tree}.}\vspace*{-1.2em}
		\label{fig:egrk}
	\end{figure*}
		Let $V=\Set{1,2,3,4}$, and 
		\begin{align*}
		\RZ_1=\RX_a, \ \RZ_2=(\RX_a,\RX_b), \ \RZ_3=(\RX_b,\RX_c),  \text{ and }   \RZ_4=\RX_c
		\end{align*}
		where $\RX_i$'s are independent uniformly random
		bits. The optimal partitions in $\Uppi^*(\RZ_V)$ are shown in
		\figref{fig:egrk:optpart}, and $\pzT_{\max}(\RZ_V)=\Set{\Set{1,2,3},\Set{2,3,4}}$. 
		Although $\pzT_{\max}(\RZ_V)$ is not an optimal partition,
		$\bar{\pzT}_{\max}(\RZ_V)=\Set{\Set{4},\Set{1}}$, which
		satisfies~\eqref{eq:T2}. By \eqref{eq:C2}, the set of minimum
		critical edges is $\pzS_{\op{crit}}(\RZ_V)=\Set{\Set{1,4}}$. It crosses at least two blocks of every coarsest
		optimal partition, and therefore every optimal partition.
		
		Example~\ref{eg:tree} is a special case of the pairwise independent network (PIN) source model
		\cite{nitinawarat-ye10}, where in this example the source is simply a tree, see \figref{fig:egrk:source}. For tree PINs in general, it can be argued that a partition $\mcP\in \Pi'(V)$ is optimal iff each block $C\in \mcP$ induces a tree on $C$. For example,
		the optimal partition $\Set{\Set{1,2},\Set{3,4}}$ in
		\figref{fig:egrk:optpart} induces two
		subtrees, one connecting $1$ and $2$, and the other connecting $3$
		and $4$. Thus, the
		fundamental partition is always the partition into singletons, but it
		is not the only optimal partition for $\abs{V}\geq 3$. Nevertheless, we can show that there is only
		one critical edge and such an edge is equal to the set of leaves.
\end{Example}

\section{Computing the growth rate of different orders}

In this section, we will illustrates the computation of $`r_k^+$ in \eqref{eq:`r_k^+}.
There are simple cases where $`r_k^+$ (and therefore $\pzS_{\op {crit}}$) can be computed
easily. For example, in the special case when the fundamental partition is the unique
optimal partition,\mnote{a:un} i.e.\ $\abs{\Uppi^*(\RZ_V)}=1$, it can be argued easily from
\eqref{eq:rhoS} that 
\begin{align*}
	`r_k^+(\RZ_V)=\frac{k-1}{\abs{\pzP^*(\RZ_V)}-1} \kern1em \text{for
		$k\leq |\pzP^*(\RZ_V)|$},
\end{align*}
where an optimal solution $S$ to \eqref{eq:`r_k^+} is any set of $k$
elements, each from a different block of the fundamental
partition. In particular, 
\begin{align*}
	\pzS_{\op {crit}}(\RZ_V) = \Set{\Set{i,j}\mid i\in C, j\in V`/C, C\in \pzP^*(\RZ_V)}
\end{align*}
and so all the critical edges are minimum with size $2$. When the
partition into singletons is the unique optimal partition, then the
optimal solution $S$ to \eqref{eq:`r_k^+} is simply any $k$-subset of $V$.
In particular, the singleton partition is shown to be the unique optimal partition in \cite{mukherjee14} for any PIN model
that corresponds to a complete graph. We can also show the same result for any PIN model that corresponds to a cycle. In general, it is possible to
check in strongly
polynomial time whether the fundamental partition is the unique optimal partition. (See Appendix~\ref{sec:uniqueness}.) For instance, in Example~\ref{eg:patch}, the optimal partition is unique and all the critical edges have size $2$ as expected.
More details about the computation and interpretations of the fundamental partition can be found in \cite{chan15mi}.

When the optimal partition is not
unique, $`r_k^+(\RZ_V)$ may not be
linear in $k$. The marginal increase in growth rate may not be diminishing in $k$ either. This is
the case, for instance, for Example~\ref{eg:tree} as shown in \figref{fig:egrk:`r_k}. 

If $H(\RZ_B)$ is an integer for every $B\subseteq V$, then
$`r_S^+(\RZ_V)$ can be computed in strongly polynomial time. To argue
this, choose $`e=\frac1{\abs{V}!}$. Note that $I_{\mcP}(\RZ_V)/`e$ is
an integer and so $I_{\mcP}(\RZ_V)$ for different $\mcP\in \Pi'(V)`/\Uppi^*(\RZ_V)$
is larger than $I(\RZ_V)$ by at least $`e$, while
$I(\RZ_V^{S,`e})$ is larger than $I(\RZ_V)$ by at most
$`e$. Thus, $\Uppi^*(\RZ_V^{S,`e})\subseteq \Uppi^*(\RZ_V)$ and so
	$`r_S^+(\RZ_V)= \frac{I(\RZ_V^{S,`e})-I(\RZ_V)}{`e}$.
Each term on the R.H.S.\ can be computed in strongly polynomial
time. A more refined argument suggests that one can choose any $`e\leq
\frac1{(\abs{V}-1)(\abs{V}-2)}$.

\section{Conclusion}

In this work, we have formulated the ISKA and DSKA problems to study how the MMI of a set of random variables changes with respect to an infinitesimal increment and decrement in the MMI of a subset of the random variables.
We found that the set of opimal partitions that attain the MMI of a set of random variables completely characterizes the growth rate for ISKA and the loss rate for DSKA. 

For ISKA, the growth rate can be computed easily in some special cases, e.g., when the optimal partition is unique. In general, however, it is not clear whether the computation can be done in polynomial time. The growth rate can be non-linear in the order, and the marginal return may even increase as we increase the order. Very surprisingly, however, a minimum critical edge can be computed in strongly polynomial time because all critical edges have the same size. In other words, one can easily identify a minimum subset of users to give an additional common randomness to, such that the secrecy capacity strictly increases. 
For DSKA, the condition for an edge to be redundant can be characterized in strongly polynomial time using the fundamental partition. Identifying excess edges is useful in simplifying secret key agreement schemes. In particular, it is hopeful that further investigation can resolve the conjectures regarding the communication complexity for secret key agreement~\cite{MKS16,mukherjee16,chan16itw}.


\appendices

\makeatletter
\@addtoreset{equation}{section}
\renewcommand{\theequation}{\thesection.\arabic{equation}}
\@addtoreset{Theorem}{section}
\renewcommand{\theTheorem}{\thesection.\arabic{Theorem}}
\@addtoreset{Lemma}{section}
\renewcommand{\theLemma}{\thesection.\arabic{Lemma}}
\@addtoreset{Corollary}{section}
\renewcommand{\theCorollary}{\thesection.\arabic{Corollary}}
\@addtoreset{Example}{section}
\renewcommand{\theExample}{\thesection.\arabic{Example}}
\@addtoreset{Remark}{section}
\renewcommand{\theRemark}{\thesection.\arabic{Remark}}
\@addtoreset{Proposition}{section}
\renewcommand{\theProposition}{\thesection.\arabic{Proposition}}
\@addtoreset{Definition}{section}
\renewcommand{\theDefinition}{\thesection.\arabic{Definition}}
\@addtoreset{Subclaim}{Theorem}
\renewcommand{\theSubclaim}{\theLemma.\arabic{Subclaim}}
\makeatother

\section{Proofs of basic properties}
\label{sec:A}

\begin{Proof}[Theorem~\ref{thm:rhopatch}]
	Clearly, $`r_{`0}^+(\RZ_V)=0$. Consider $S\subseteq V:S\neq `0$. Rewriting the divergence in terms of the entropy as $D(P_{\RZ_V}\|\prod_{C\in \mcP}P_{\RZ_C})=\sum_{C\in \mcP} H(\RZ_C)-H(\RZ_V)$, we have
	\begin{align*}
		I(\RZ^{`e,S}_V) &= \min_{\mcP\in \Pi'(V)}
		\frac1{\abs{\mcP}-1} `1[\sum_{C\in \mcP} H(\RZ^{`e,S}_C) -
		H(\RZ^{`e,S}_V) `2]\\
		&= \min_{\mcP\in \Pi'(V)}
		\frac1{\abs{\mcP}-1} \Bigg[\sum_{C\in \mcP}
		H(\RZ_C)-H(\RZ_V)\\
		&\kern1em +\sum_{C\in \mcP} H(\RX)`c_{\Set{S\cap C\neq `0} }
		- H(\RX) \Bigg]\\
		&= \min_{\mcP\in \Pi'(V)} `1[I_{\mcP}(\RZ_V) + `e \frac{\sum_{C\in \mcP} `c_{\Set{C \cap S\neq `0}} - 1}{\abs{\mcP}-1}`2].
	\end{align*}
	Suppose $I(\RZ^{`e,S}_V)< \min_{\mcP\in \Pi'(V)`/\Uppi^*(\RZ_V)}
	I_{\mcP}(\RZ_V)$, which is possible for all $`e>0$ sufficiently
	small since
	$\abs{\Pi'(V)}$ is finite.
	It does not lose optimality to restrict $\mcP$ to $\Uppi^*(\RZ_V)$ and so
	we have for all $`e>0$ sufficiently small that
	\begin{align*}
		I(\RZ^{`e,S}_V) &= I(\RZ_V) + `e \min_{\mcP\in
			\Uppi^*(\RZ_V)}  \frac{\sum_{C\in \mcP^*} `c_{\Set{C \cap S\neq `0}} - 1}{\abs{\mcP}-1}
	\end{align*}
	which gives \eqref{eq:rhoS}. \eqref{eq:patch} follows from
	\eqref{eq:rhoS} directly.
\end{Proof}

\begin{Proof}[Theorem~\ref{thm:rho-}]
	The proof of \eqref{eq:rhoS-} is analogous to the proof of \eqref{eq:rhoS} above, but we have $\max$ instead of $\min$ due to a sign change because the rate is on the loss rather than growth of the MMI. To derive the condition~\eqref{eq:excess} for excess edge, notice that \eqref{eq:rhoS-} is zero iff $S$ does not cross any optimal partitions. It suffices to consider only the finest optimal partition, namely $\pzP^*(\RZ_V)$, because $S$ does not cross $\pzP^*(\RZ_V)$ implies it does not cross any coarser optimal partitions, which cover all the optimal partitions.
\end{Proof}

\begin{Proof}[Proposition~\ref{pro:rhopatch}]
	$`r_k^+(\RZ_V)$ is non-decreasing in $k$ because $`r_S^+(\RZ_V)$ by \eqref{eq:rhoS} is
	non-decreasing in $S$ with respect to set inclusion. It is equal to
	$0$ for $k=1$ because $`r_{\Set{i}}^+(\RZ_V)=0$ for all $i\in V$. This also means that a critical edge, if any, must be non-singleton, with size at least two. $`r_k^+(\RZ_V)$ is at most
	$1$ because $`r_V^+(\RZ_V)=1$. More precisely, $`r_S^+(\RZ_V)=1$ iff $S\cap C\neq `0$ for every $C\in \mcP^*$ and $\mcP^*\in
	\Uppi^*(\RZ_V)$. Since the fundamental partition $\pzP^*(\RZ_V)$ is
	the unique finest optimal partition, we have $`r_S^+(\RZ_V)=1$ iff $S\cap C\neq `0$ for every $C\in \pzP^*(\RZ_V)$. That means
	$`r_S^+(\RZ_V)<1$ if $\abs{S}<abs {\pzP^*(\RZ_V)}$, and $`r_S^+(\RZ_V)=1$ for
	any $S$ obtained by taking at least one element from each block in the
	fundamental partition. The fact that $`r_V^+(\RZ_V)=1$ also means that there is at least one critical edge. 
\end{Proof}


\section{Proof of Theorem~\ref{thm:TC}}
\label{sec:B}

\begin{Proof}[Theorem~\ref{thm:TC}]
	From \eqref{eq:patch}, we have
	$S\in \pzS_{\op {crit}}(\RZ_V)$ iff 
	\begin{align}
		S\nsubseteq C\kern1em\text{or equiv.}\kern1em S`/C\neq `0 \kern1em
		\forall C\in \pzT_{\max}(\RZ_V).\label{eq:TC}
	\end{align}
	From this, it can
	be argued easily that the sets defined in \eqref{eq:C1} and \eqref{eq:C2}
	are critical edges for the cases \eqref{eq:T1} and \eqref{eq:T2}
	respectively. If $S$ is a critical edge under \eqref{eq:T1}, any
	element, say $i\in S$,  must be contained by some $C\in
	\pzT_{\max}(\RZ_V)$ since $\pzT_{\max}(\RZ_V)$ is a partition of $V$. By
	\eqref{eq:TC}, it must contain an element $j\in V`/C$ and so $S$
	must be in \eqref{eq:C1} as desired. If $S$ is a critical edge under
	\eqref{eq:T2}, it must contain an element from $V`/C$ for every
	$C\in \pzT_{\max}(\RZ_V)$ by \eqref{eq:TC}. Thus, it must be in
	\eqref{eq:C2} as desired.
\end{Proof}

It remains to prove Lemma~\ref{lem:T}. We do so using the idea of
\emph{zero-singleton-submodular} function in \cite{narayanan90}. Denote
the fundamental partition as
\begin{align}
	\pzP^*(\RZ_V) = (C^*_1,\dots,C^*_{\ell})
\end{align}
by indexing the blocks from $1$ to $\ell=\abs{\pzP^*(\RZ_V)}$. Define
$g:2^{[\ell]}\mapsto `R$ as
\begin{align}
	g(B) &:= h_{`g}`1(\bigcup\nolimits_{i\in B} C^*_i`2) - \sum_{i\in B}  h_{`g}(C^*_i)
	\kern1em\text{for $B\subseteq [\ell]$}\kern-.2em\label{eq:zss:g}
\end{align}
with $`g:=I(RZ_V)$ and $h_{`g}(C):=H(\RZ_C)-`g$ is the residual randomness defined in \cite{chan15mi}. 
It follows immediately that $g$ is submodular with
\begin{align}
	g(\Set{i})=0\kern1em \text{for $i\in [\ell]$,}\label{eq:zs}
\end{align}
and is therefore called a
zero-singleton-submodular function. It can also be shown to be
non-negative over non-empty sets.
\begin{Proposition}[\mbox{\cite[p.198]{narayanan90}}]
	\label{pro:zss}
	$g(B)\geq 0$ for all $B\subseteq [\ell]:B\neq `0$. The
	\emph{zero sets} of $g$,
	\begin{align}
		\pzZ(g):=\Set{B\subseteq [\ell]:g(B)=0}\label{eq:zsets}
	\end{align}
	forms an \emph{intersecting family}, i.e.\ 
	\begin{align}
		U\cap W,U\cup W\in \pzZ(g)\text{ for all }U,W\in \pzZ(g):U\cap W\neq `0
	\end{align}
	Furthermore,
	\begin{align}
		\Set*{\bigcup\nolimits_{i\in B}C^*_i:B\in
			\pzZ(g)}=\bigcup\Uppi^*(\RZ_V) \cup\Set{V},\label{eq:zsets->P}
	\end{align}
	and so $\bigcup\Uppi^*(\RZ_V) \cup\Set{V}$ is also an intersecting family.
\end{Proposition}
\begin{Proof}
	For any partition $\mcP\in
	\Pi([\ell])$,
	\begin{align*}
		g[\mcP]&:=\sum_{C\in \mcP} g(C)\\
		&= \sum_{C\in \mcP} h_{`g}`1(\bigcup\nolimits_{i\in C} C^*_i`2) - \sum_{i=1}^\ell
		h_{`g}(C^*_i)\\
		&= h_{`g}`1[\Set*{\bigcup\nolimits_{i\in C} C^*_i:C\in \mcP}`2] -
		h_{`g}[\pzP^*(\RZ_V)]\geq 0,
	\end{align*}
	because, by \cite[Theorem~5.1]{chan15mi}, $\pzP^*(\RZ_V)$ minimizes
	$h_{`g}$ over all partitions of $V$, which include $\Set{\bigcup_{i\in C} C^*_i:C\in
		\mcP}$. Furthermore, we have equality $g[\mcP]=0$ for $\mcP\in \Pi'([\ell])$ iff  
	\begin{align*}
		\Set*{\bigcup\nolimits_{i\in C} C^*_i:C\in \mcP} \in \Uppi^*(\RZ_V).
	\end{align*}
	Suppose to
	the contrary that $g(B)<0$ for some non-empty $B\subseteq [\ell]$. Then,
	$g[\Set{B}\cup\Set{\Set{i},i\in [\ell]`/B}]<0$ by the zero-singleton
	property~\eqref{eq:zs}, but this contradicts $g[\mcP]\geq 0$ above
	for all $\mcP\in \Pi([\ell])$. This proves the non-negativity of $g$
	over non-empty sets.
	
	For $U,W\in \pzZ(g):U\cap W\neq `0$, we have
	\begin{align*}
		0\geq g(U)-g(U\cap W) \geq g(U\cup W) - g(W) \geq 0
	\end{align*}
	where the second inequality is by the submodularity of $g$, and the
	first and last inequalities are because $g(U)=g(W)=0$ and $g(U\cap
	W),g(U\cup W)\geq 0$ by the non-negativity of $g$ over non-empty
	sets argued above. Thus, all inequalities are satisfied with
	equality and so $g(U\cap W)=g(U\cup W)=0$ as desired for $\pzZ(g)$
	to be an intersecting family.
	
	It remains to prove \eqref{eq:zsets->P}. By \cite[Theorem~5.2]{chan15mi}, every partition in $\Uppi^*(\RZ_V)\cup \Set{\Set{V}}$
	is coarser than $\pzP^*(\RZ_V)$ and can therefore be expressed as $\Set{\bigcup_{i\in C} C^*_i:C\in
		\mcP}\in \Pi'(V)$ for some $\mcP\in \Pi'([\ell])$. By optimality,
	$g[\mcP]=0$, and so $g(C)=0$ for all $C\in \mcP$ since $C$ is
	non-empty and $g$ is non-negative over non-empty sets as argued
	before. Conversely, suppose $g(B)=0$. Then $g[\Set{B}\cup\Set{\Set{i},i\in
		[\ell]`/B}]=0$ by \eqref{eq:zs}, and so $\bigcup_{i\in B}C^*_i$ is
	a block of an optimal partition, namely $\Set{\bigcup_{i\in
			B}C^*_i}\cup\Set{C^*_i:i\in [\ell]`/B}$. This completes the proof
	of \eqref{eq:zsets->P}. Since $C^*_i$'s are disjoint, the fact that
	$\pzZ(g)$ is an intersecting family implies that $\bigcup\Uppi^*(\RZ_V)\cup
	\Set{V}$ is.
\end{Proof}

\begin{Proof}[Lemma~\ref{lem:T}]
	We first argue that: for any distinct $C_1,C_2\in
	\pzT_{\max}(\RZ_V)$, we have 
	\begin{align}
		C_1\cap C_2\neq `0 \text{ implies }C_1\cup C_2= V.\label{eq:lt1}
	\end{align}
	To show this, note that $C_1$ and $C_2$ are maximal sets from
	$\bigcup \Uppi^*(\RZ_V)$ by definition~\eqref{eq:T}. By
	Proposition~\ref{pro:zss}, $\bigcup\Uppi^*(\RZ_V)\cup \Set{V}$ is an
	intersecting family and so $C_1\cap C_2\neq `0$ implies $C_1\cup
	C_2$ is also in the family. Suppose to the contrary that $C_1\cup
	C_2\neq V$, then $C_1\cup C_2$ is a strictly larger set in
	$\bigcup\Uppi^*(\RZ_V)$ than the distinct sets $C_1$ and $C_2$, which
	contradicts the maximality. 
	
	Next, we argue that if there exists distinct $C_1,C_2\in
	\pzT_{\max}(\RZ_V)$ such that $C_1\cap C_2\neq `0$, then 
	\begin{align}
		C\cap C'\neq `0\text{ for all }C,C'\in \pzT_{\max}(\RZ_V).\label{eq:lt2}
	\end{align}
	Indeed, we have a stronger statement that
	\begin{align}
		`0\neq (V`/C_1)\subseteq C\text{ for all }C\in
		\pzT_{\max}(\RZ_V)`/\Set{C_1}.
		\label{eq:lt3}
	\end{align}
	If $C=C_2$, we have \eqref{eq:lt3} directly from \eqref{eq:lt1}. Suppose $C\neq C_2$. Then,
	$C_1\cap
	C\neq `0$ because $C\cap V`/C_2\neq `0$ by the maximality of $C\neq
	C_2$, and
	$V`/C_2 \subseteq C_1$ by \eqref{eq:lt1} that $C_1\cup
	C_2=V$. Applying \eqref{eq:lt1} again with $C_2$ replaced by $C$, we
	have $C_1\cup C=V$ as desired.
	
	By the contrapositive statement of \eqref{eq:lt3}, if $C_1\cap
	C_2=`0$ for some distinct $C_1,C_2\in \pzT_{\max}(\RZ_V)$, then all
	the sets in $\pzT_{\max}(\RZ_V)$ must be disjoint, and so we have
	\eqref{eq:T1} since every element in $V$ must be covered by at least
	one block of an optimal partition. In the other case when every
	distinct $C_1,C_2\in \pzT_{\max}(\RZ_V)$ must intersect,
	\eqref{eq:lt1} implies
	$C_1\cup C_2=V$, or equivalently, $(V`/C_1)\cap (V`/C_2)=`0$, which
	implies \eqref{eq:T2} as desired.
\end{Proof}

By \eqref{eq:zsets->P}, $\pzT_{\max}(\RZ_V)$ can be obtained from the maximal zero sets in
$\pzZ(g)$ together with the fundamental
partitions $\pzP^*(\RZ_V)$. The maximal zero set not containing an element
$i\in [\ell]$, can be computed
in strongly polynomial time as follows: Starting with $C=`0$, add an
element from $[\ell]`/\Set{i}$ to $C$
repeatedly as long as $\min_{C'\in 2^{[\ell]`/\Set{i}}: C\subseteq
	C'} g(C')=0$. The minimization is repeated at most $O(\abs{V}^2)$
times and can be solved in strongly
polynomial time using some existing algorithm for submodular function minimization over a
lattice family~\cite{schrijver02}. Hence, $\pzT_{\max}(\RZ_V)$ and
therefore $\pzS_{\op{crit}}(\RZ_V)$ can be computed in strongly polynomial
time by Theorem~\ref{thm:TC}.

\pbox[Conjecture]{Does critical edges have the same rate of increase? We
	conjecture that $`r_S^+=\frac{\abs{S}-1}{\abs{\pzP^*(\RZ_V)}-1}$ for all $S\in
	\pzS_{\op {crit}}(\RZ_V)$.}

\section{Uniqueness of the optimal partition}
\label{sec:uniqueness}

To check whether there is a unique optimal partition in strongly polynomial time, we can make use of
the zero-singleton-submodular function $g$ defined in
\eqref{eq:zss:g} based on the fundamental partition
$\pzP^*(\RZ_V)$ and the MMI. In essense of \eqref{eq:zsets->P}, we only need
to check whether the zero sets $\pzZ(g)$ in \eqref{eq:zsets}
consists only of the singletons. This can computed in $O(\abs{V}^2)$
submodular function minimizations, namely $\min_{B\supseteq \Set{i,j}}
g(B)$ for all pair $(i,j)$ of distinct elements. $\pzP^*(\RZ_V)$
and $I(\RZ_V)$ can also be computed in $O(\abs{V}^2)$
submodular function minimizations~\cite{nagano10}.

%
%


\section*{Acknowledgments}
The authors would like to thank Prof.\ Navin Kashyap, Manuj Mukherjee, and the colleagues at INC for detailed discussion and valuable comments.

\bibliographystyle{IEEEtran}
\bibliography{IEEEabrv,ref}

\begin{IEEEbiography}[{\includegraphics[width=1in,height=1.25in,clip,keepaspectratio]{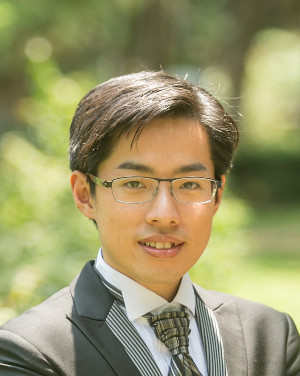}}]{Chung Chan} received the B.Sc., M.Eng. and Ph.D. from the EECS Department at MIT in 2004, 2005 and 2010 respectively. He is currently a Research Assistant Professor at the Institute of Network Coding, the Chinese University of Hong Kong. 	
	His research is in the area of information theory, with applications to network coding, multiple-terminal source coding and security problems that involve high-dimensional statistics. He is currently working on machine learning applications such as data clustering and feature selection. 
\end{IEEEbiography}
\vspace*{-2em}
\begin{IEEEbiography}[{\includegraphics[width=1in,height=1.25in,clip,keepaspectratio]{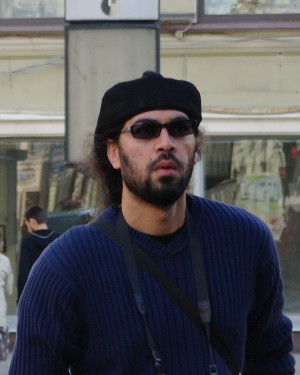}}]{Ali Al-Bashabsheh}
	received a B.Sc. (2001) and an M.Sc (2005) in electrical engineering from Jordan University of
	Science and Technology,
	an M.Sc. (2012) in mathematics from Carleton University,
	and a Ph.D. (2014) in electrical engineering from the University of Ottawa. 
	Since April 2014, he has been a postdoctoral fellow at the Institute of Network Coding
	at the Chinese University of Hong Kong. His research interests include graphical models, coding
	theory, and information theory.
\end{IEEEbiography}
\vspace*{-2em}
\begin{IEEEbiography}[{\includegraphics[width=1in,height=1.25in,clip,keepaspectratio]{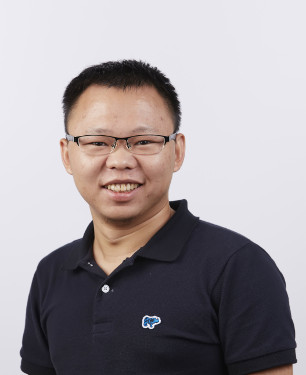}}]{Qiaoqiao Zhou} received his B.B.A. in business administration and M.S. in electrical engineering from Beijing University of Post and Telecommunication, China, in 2011 and 2014, respectively. In 2014-2015 he was a research assistant at Institute of Network Coding, the Chinese University of Hong Kong. He is currently a Ph.D. student at Department of Information Engineering, the Chinese University of Hong Kong. His research interests include information-theoretic security and machine learning. 
\end{IEEEbiography}

\end{document}